\documentclass[aps,nofootinbib]{revtex4-2}
\usepackage{epsfig}
\usepackage{epsf}

\begin{document}

\title{Correlations equalities and  some upper bounds for the coupling constant  implying area decay of Wilson loop for $Z_3$ lattice gauge theories}

\author{A. L. Mota}
\email{motaal@ufsj.edu.br}
\affiliation{Departamento de Ci\^{e}ncias Naturais, Universidade Federal de S\~{a}o Jo\~{a}o del Rei, \\
C.P. 110,  CEP 36301-160, S\~ao Jo\~ao del Rei, Brazil}
\author{F. C. S\'{a} Barreto}
\email{fcsabarreto@gmail.com}
\affiliation{Departamento de F\'{i}sica, Universidade Federal de Minas Gerais, C.P. 110, CEP 31270-901, Belo Horizonte, MG, Brazil}

\begin{abstract}
Correlation identities are obtained for $Z_3$ lattice gauge theory where the bonds of the plaquettes are decorated by generalized three-state Ising variables. Making use of correlation inequalities we obtain the area decay of the Wilson loop observable in a range of the coupling parameter larger than those obtained from mean field theory considerations.
\end{abstract}

\maketitle

\section{Introduction}

Correlation inequalities combined with exact identities are useful in obtaining rigorous results in statistical mechanics. Among the various questions that are resolved by them one is the decay of the correlation functions. The decay of the correlation functions give information about the critical couplings of the statistical mechanics model. In this publication, the method will be applied to study the $Z_3$ lattice gauge problem where the bond variables are decorated by generalized three-state Ising variables $e^{n 2\pi i/3}$, n=0,1,2. The $Z_3$ discrete gauge symmetry is relevant to many fields of physics, such as, for instance, in the study of confinement/deconfinement transition in particle physics \cite{Jiahui1999,Anber2013,Robaina2021,Baranka2022,Akiyama2023,Celik1985}, in dark matter models, where the $Z_3$ symmetry ensures stability of dark matter particles \cite{Ko2015,Choi2015,Choi2020,Borah2023}, and, of course, in the study of lattice gauge theories and their transitions \cite{Drouffe1981,Emonts2020,Gattringer2012,Fisch1994,Brower1991,Hattori2010}. 

In this work, we firstly obtain correlation identities for the  $Z_3$ gauge lattice, which are the gauge version of the spin identities obtained previously for the spin one Blume-Capel model (\cite{Brydges1982,Siqueira1986,SaBarreto2012,SaBarreto2015}). We then look for the area decay of Wilson's loop, which is a criterion for confinement. We apply  the first and second Griffiths inequalities \cite{Griffiths1969,Ginibre1969,Sylvester1976,Simon1980,Ganikhodjaev2007,Ganikhodjaev2010} and Newman's inequalities \cite{Newman1975, Sylvester1976} to the correlation identities to obtain the coupling parameter which gives the area decay of the Wilson loop in d=3 and d=4 dimensions. In both dimensions the coupling parameters obtained for the area decay of the Wilson loop improve over the mean field results.

We consider the well known Wilson loop observable of a $ Z_3 $  pure gauge $ Z^d $ lattice theory with Wilson action defined as
\begin{equation}
\langle W( C )\rangle  = \lim_{\Lambda \to Z^d} \langle  W( C) \rangle _\Lambda \label{eq1}  
\end{equation}               
where $\langle W(C)\rangle $ is the finite lattice Gibbs ensemble average.
The Wilson action Boltzmann factor \cite{Wilson1974,Seiler1982,Kogut1979},is given by,
\begin{equation}
e^{{\beta J\sum_{P\subset\Lambda}\chi_P}  } \label{eq2}
\end{equation}                
where P denotes the unit squares (plaquettes) of $\Lambda$ and $\chi _P$ = $S_1S_2S_3S_4$. 
We let $S_b$ denote the bond variables which take values $e^{0}=1, e^{2\pi i/3}, e^{4\pi i/3}$ . W(C) is the product of $S_b$ along the perimeter of the planar rectangle C of area A. The gauge coupling constant is $\beta  J$ , $0 <  \beta  J < \infty$. 
Area decay of $\langle W( C )\rangle $ is a criterion for confinement. For the $Z_2$ lattice gauge theory it is known that area decay occurs for small $\beta J$ and for sufficiently large $\beta J$ perimeter decay occurs for $d \geq  3$ \cite{Seiler1982, Kogut1979}. For $d = 2$, $\langle W( C )\rangle $  has area decay for all $\beta J$ by explicit calculation \cite{Kogut1979}. We take free boundary conditions and note that Griffiths first and second inequalities  apply (\cite{Szasz1978, Griffiths1967a,Griffiths1967b,Griffiths1967c}), and especially for the Potts model see (\cite{Ganikhodjaev2007,Ganikhodjaev2010}), and therefore imply the existence of the thermodynamic limit \cite{Glimm1981}.
For d = 3 and 4 we obtain lower bounds $\beta_L$  on the area decay of $\langle W( C )\rangle $ i,.e. for all $\beta < \beta_L$, $\langle W ( C )\rangle $ has area decay, using correlation identities and Griffiths inequalities.  The correlation identities are a gauge version of the identities  presented before \cite{Siqueira1986,SaBarreto2012,SaBarreto2015} and employed to obtain lower than mean field upper bounds on the critical temperature for the Blume - Capel spin 1 systems. The procedure is similar to that employed previously \cite{SaBarreto1983a} to obtain the $Z_2$ lattice gauge correlation identities as a version of Callen's identities for the Ising spin 1/2 systems. 
For completeness, in section \ref{section2}, we give the mean field lower bound $\beta _M$ $(\beta_M < \beta_L)$ using a decoupling and Griffiths inequality argument \cite{SaBarreto1983b, Tomboulis1981}.

\section{Mean field lower bounds} \label{section2}
For definiteness assume that C lies in the ($x_1$, $x_2$)  plane. Consider a bond b fixed in the lower left-hand corner of C. Replace $\beta$  by $\beta \lambda$, $\lambda  \epsilon  [0,1]$, in the action for the $2(d-1)$ plaquettes (call them $P_1,P_2,...,P_{2(d-1)})$ that have one bond in common with b. Denote the corresponding expectation by $\langle W(C)\rangle _\lambda$, and let $J=1$, without loss of generality. Integrating $(d/d\lambda ) \langle W(C)\rangle _\lambda $ gives, noting that $\langle W(C)\rangle _0 = 0$,
\begin{eqnarray}
&& \langle W(C)\rangle  =\int_{0}^{1} d\lambda \frac{d}{d\lambda } \langle W(C)\rangle _\lambda = \beta \int_{0}^{1} d\lambda \sum_{i=1}^{2(d-1)} \left (\langle W(C) \chi _{P_i}\rangle _\lambda - \langle W(C)\rangle _\lambda \langle  \chi _{P_i}\rangle _\lambda \right)\nonumber \\
&&\leq \beta  \int_{0}^{1} d\lambda \sum_{i=1}^{2(d-1)} \langle W(C) \chi _{P_i}\rangle _\lambda \leq \beta \sum_{i=1}^{2(d-1)} \langle W(C) \chi _{P_i}\rangle ,\label{eq3}            
\end{eqnarray}             
In the previous we used Griffith's first (second) inequality in the first (second) equality. Each term on the right corresponds to a modified contour determined by the bonds of the variables of  $W( C ) \chi_{P_i}$ which enlarges or diminishes C by one plaquette. We repeat the argument proceeding along successive rows of plaquettes enclosed by C. After A applications we arrive at
\begin{equation}
\langle W( C )\rangle  \leq \beta^A \text{(sum of } [2(d-1)]^A \text{ terms)}         \label{eq4}   
\end{equation} 
Each term is non-negative and bounded above by 1 giving $\langle W( C )\rangle  \leq  [\beta  2(d-1)]^A$.
Therefore, for each $\beta$, 
\begin{eqnarray}
&&\beta \subset (0,(2(d-1)) ^ -1),\nonumber\\
&& \langle W( C )\rangle  \leq  e^{-\mid \ln (2\beta (d-1))\mid A}.     \label{eq5}  
\end{eqnarray} 

\section{ Correlation Identities for the $Z_3$ Lattice Gauge Model }

We define the action for the $Z_3$  lattice gauge model as
\begin{equation}
-H = \beta J\sum_{P\subset \Lambda }\chi _P  \label{eq21}
\end{equation}
where, $\chi _P$ is the square plaquette and $S_i$ is the bond variable assuming values  $e^{0}=1, e^{2\pi i/3} \equiv \sigma$, and $e^{4\pi i/3} = \sigma^2$.
Let $S_D = S_{i_1}...S_{i_D}$ denote a product of distinct bond variables and for a fixed bond b occurring in $S_D$ give a numerical ordering 1,2,... to the 2(d- 1) plaquettes that have one bond in common with b. 
We have,
\begin{equation}
\Big\langle S_D\Big\rangle =\frac{1}{Z} \sum_{S}S_D e^{\beta J\sum_{P\subset \Lambda }\chi _P }
\end{equation}
where,
\begin{equation}
Z= \sum_{{S}} e^{\beta J\sum_{P\subset \Lambda }\chi _P }
\end{equation}
Let us consider the bond b, with $S_b$, and the plaquette $\chi _b$ which contains the bond b.
Let $S_D^{(b)}$ be the product $S_D$ with the bond b deleted. We have,
\begin{equation}
\Big\langle S_D\Big\rangle =\frac{1}{Z}\cdot  \Big[\sum_{S}S_D^{(b)}\Big(\frac{\sum_{S_b}S_b e^{\beta J\chi _b}}{\sum_{S_b}e^{\beta J\chi _b }}\Big)\cdot e^{\beta J\sum_{P\subset \Lambda }\chi _P }\Big]\label{eq22}
\end{equation}
Let
\begin{equation}
\chi _b = S_b\sum_{neig.(b)}S_kS_lS_m
\end{equation}
Summing over $S_b$, in (\ref{eq22}) and introducing $\nabla$ = $\frac{\partial}{\partial x}$ through $e^ {a\nabla } f(x)=f(x+a)$, we get
\begin{equation}
\left\langle S_D\right\rangle =\left\langle S_D^{(b)}\left(\frac{e^{\alpha_b}+2 e^{-\frac{1}{2}\alpha_b} \cos\left( \frac{\sqrt{3}}{2} \alpha_b + \frac{2 \pi}{3} \right)}{e^{\alpha_b}+2 e^{-\frac{1}{2}\alpha_b} \cos\left( \frac{\sqrt{3}}{2} \alpha_b \right)}\right)\right\rangle
 = \left\langle S_D^{(b)}\prod_{neig(b)}e^{(\beta J S_k S_l S_m) \nabla} \right\rangle \cdot f(x)\mid_{x=0} \label{eq24} 
\end{equation}
where,  
\begin{equation}
\alpha_b = \beta J\sum_{neig.(b)} S_k S_l S_m
\end{equation}
and
\begin{equation}
f(x) = \frac{e^x + \sigma e^{\sigma x} + \sigma^2 e^{\sigma^2 x}}{e^x + e^{\sigma x} + e^{\sigma^2 x}} = \frac{e^x+2e^{-\frac{1}{2}x} \cos\left( \frac{\sqrt{3}}{2} x + \frac{2\pi}{3} \right)}{e^x+2e^{-\frac{1}{2}x} \cos\left( \frac{\sqrt{3}}{2} x \right)}.  \label{eqfx}
\end{equation}

Expanding the exponential in (\ref{eq24}) and using $S_i^{3n} = S_i^{3}=1$, $S_i^{3n+1} = S_i$, and $S_i^{3n+2} = S^2_i$, for $n=1,2,...$, we obtain,
\begin{equation}
\Big\langle S_D\Big\rangle =\Big\langle S_D^{(b)}\prod_{neig(b)}\Big[c_0(\beta J\nabla) + S_k S_l S_m s_1(\beta J\nabla ) + S^2_k S^2_l S^2_m s_2(\beta J\nabla ) \Big]
\Big\rangle \cdot f(x)\mid_{x=0}, \label{eq26} 
\end{equation}
with
\begin{equation}
c_0(x) = \frac{e^x+e^{\sigma x}+e^{\sigma^2 x}}{3}; \textrm{ }s_1(x) = \frac{e^x+\sigma^2 e^{\sigma x}+\sigma e^{\sigma^2 x}}{3}; \textrm{ and } s_2(x) = \frac{e^x+\sigma e^{\sigma x}+\sigma^2 e^{\sigma^2 x}}{3}. \label{c0s1s2}
\end{equation}

The functions $c_0(x), s_1(x)$, and $s_2(x)$ were named by their similarity to the hyperbolic $\cosh(x)$ and $\sinh(x)$ that appear in the equivalent treatment of the $Z_2$ lattice gauge \cite{SaBarreto1983a}. The following properties, derived from Eqs. (\ref{eqfx}) and (\ref{c0s1s2}) , are usefull in the evaluation of $\langle S_D \rangle$ in the next section:
\begin{equation}
f(\sigma x) = \sigma^2 f(x), \textrm{ } f(\sigma^2 x) = \sigma f(x),
\end{equation}
\begin{equation}
1 + \sigma + \sigma^2 = 0, \textrm{ } \sigma^3 = 1,
\end{equation}
and 
\begin{equation}
c^a_0(\beta J\nabla) s^b_1(\beta J\nabla) s^c_2(\beta J\nabla) f(x) \mid_{x=0} \Big\{ \begin{array}{c}\neq 0 \textrm{ if } (b+2c+1) = 3n  \\ = 0 \textrm{ otherwise},  \end{array}
\end{equation}
for $a, b, c,$ and $n$ positive integers.

In contrast with the $Z_2$ case \cite{SaBarreto1983a}, where it is given by the analytical function $tanh(x)$, $f(x)$ (Eq. (\ref{eqfx})) is non-analytical for $e^x+2e^{-\frac{1}{2}x} \cos\left( \frac{\sqrt{3}}{2} x \right) = 0$, i.e., at $x \approx -1.869812799$. However, as we shall see, this point is above both $d=3$ and $d=4$ upper limits for $\beta J$.

\section{ Exact correlation identities applied to the $d=3$, and $d=4$ lattices (square plaquettes)}

Applying Eq. (\ref{eq26}), for $d=3$ and $d=4$, after some algebra, we obtain, in subsections (\ref{ssd3}) and (\ref{ssd4}), with $k = \beta J$, $\rho_2 = 2 + \sigma$, $\rho_3 = 3 + \sigma$, $\rho_5 = 5 + \sigma$, and $\Re\{z\}$ stands for the real part of the complex argument $z$:

\subsection{d=3}\label{ssd3}
\begin{eqnarray}
\langle S_D\rangle &=&
A_1 \sum_{i} \langle S_D \chi^2_i \rangle 
+ A_2 \sum_{i<j} \langle S_D \chi_i \chi_j \rangle 
+ A_3 \sum_{i<j<k} \langle S_D \chi_i \chi^2_j \chi^2_k \rangle \nonumber \\
&+& A_4 \sum_{i<j<k<l} \langle S_D \chi_i \chi_j \chi_k \chi^2_l \rangle  
+ A_5 \sum_{i<j<k<l}\langle S_D \chi^2_i \chi^2_j \chi^2_k \chi^2_l\rangle, \label{SD3d}
\end{eqnarray}

with 

\begin{equation}
A_1=\frac{4}{9} \left( \frac{1}{3} f(4k) + f(k) - f(-2k) + 2\Re\{\rho_3 f(\rho_3 k)\} \right) \geq 0, \label{A1d3}
\end{equation}
\begin{equation}
A_2=\frac{2}{3} \left( \frac{1}{3} f(4k) + f(-2k) - \frac{4}{3} \Re\{\sigma f(\rho_3 k)\} \right) \geq 0,
\end{equation}
\begin{equation}
A_3=\frac{4}{3} \left( \frac{1}{3} f(4k) - f(k) + \frac{2}{3} \Re\{\sigma f(\rho_3 k)\} \right) \leq 0,
\end{equation}
\begin{equation}
A_4=\frac{4}{9} \left( \frac{1}{3} f(4k) + f(k) - f(-2k) + \frac{2}{3} \Re\{\sigma \rho_3 f(\rho_3^{*} k)\} \right) \leq 0,
\end{equation}
\begin{equation}
A_5=\frac{1}{9} \left( \frac{1}{3} f(4k) + 4 f(k) + 2 f(-2k) + \frac{8}{3} \Re\{\sigma f(\rho_3 k)\} \right) \geq 0. \label{A5d3}
\end{equation}

All inequalities hold for $0 \le k \le 0.924906$, where $f(-2k)$ is analytical, except Eq. (\ref{A1d3}), that holds for $0 \le k \le 0.740736$. Above this point we get $A_1 < 0$. As we shall show, the upper bound for the coupling constant lies in the interval of validity of Eqs. (\ref{A1d3})-(\ref{A5d3}).

\subsection{d=4}\label{ssd4}

\begin{eqnarray}
\langle S_D \rangle &=&
A_1 \sum_i \langle S_D \chi_i^2    \rangle + A_2 \sum_{i < j} \langle S_D \chi_i \chi_j   \rangle + A_3 \sum_{i < j<k} \langle S_D \chi_i \chi_j^2 \chi_k^2  \rangle \nonumber \\
&+& A_4 \sum_{i < j<k<l} \langle S_D \chi_i \chi_j \chi_k \chi_l^2   \rangle + A_5 \sum_{i<j<k<l<m} \langle S_D \chi_i \chi_j \chi_k \chi_l \chi_m   \rangle + A_6 \sum_{i<j<k<l} \langle S_D \chi_i^2 \chi_j^2 \chi_k^2 \chi_l^2   \rangle \nonumber \\ 
&+& A_7 \sum_{i<j<k<l<m} \langle S_D \chi_i \chi_j \chi_k^2 \chi_l^2 \chi_m^2   \rangle + A_8 \sum_{i < j<k<l<m<n} \langle S_D \chi_i \chi_j \chi_k \chi_l  \chi_m^2 \chi_n^2 \rangle \nonumber \\
&+& A_9 \sum_{i < j<k<l<m<n} \langle S_D \chi_i \chi_j^2 \chi_k^2 \chi_l^2  \chi_m^2 \chi_n^2 \rangle, \label{SD4d}
\end{eqnarray}
where the A coefficients are given by

\begin{equation}
A_1=\frac{2}{81} \left( f(6 k) + 15 f(3 k) - 10 f(-3 k) + 2 \Re \{ 10(2+\sigma) f(\rho_2 k) + (5+\sigma) f(\rho_5 k) + 5(2+\sigma) f(2 \rho_2 k) \}  \right) \ge 0, \label{A1d4}
\end{equation}

\begin{equation}
A_2=\frac{5}{81} \left( f(6k) + 6 f(3k) + 8 f(-3k) + 2 \Re \{ -4(1+2\sigma) f(\rho_2 k) + 2(1-\sigma) f(\rho_5 k) - (2+7\sigma) f(2 \rho_2 k)\}   \right) \ge 0,
\end{equation}

\begin{equation}
A_3=\frac{20}{81} \left( f(6k) - 3 f(3k) - f(-3k) + 2\Re \{ -(7+2\sigma) f(\rho_2 k) + (2+\sigma) f(\rho_5 k) + (1+2\sigma) f(2 \rho_2 k)\}   \right) \le 0,
\end{equation}

\begin{equation}
A_4=\frac{20}{81} \left( f(6k) - 3 f(3k) - f(-3k) + 2\Re \{ (5+7\sigma) f(\rho_2 k) - (1+2\sigma) f(\rho_5 k) - (2+\sigma) f(2\rho_2 k)  \}   \right) \le 0,
\end{equation}

\begin{equation}
A_5=\frac{2}{81} \left( f(6k) + 15 f(3k) - 10 f(-3k) + 2\Re \{ -10(1+2\sigma) f(\rho_2 k) - (4+5\sigma) f(\rho_5 k) - 5 (1-\sigma) f(2 \rho_2 k) \} \right) \le 0,
\end{equation}

\begin{equation}
A_6=\frac{5}{81} \left( f(6k) +6 f(3k) +8 f(-3k) + 2\Re \{ (8+4\sigma) f(\rho_2 k) + (2+4\sigma) f(\rho_5 k) + (2\sigma-5) f(2 \rho_2 k) \} \right) \ge 0,
\end{equation}

\begin{equation}
A_7=\frac{20}{81} \left( f(6k) -3 f(3k) - f(-3k) + 2\Re \{ (2-5\sigma) f(\rho_2 k) +(\sigma-1) f(\rho_5 k) + (1-\sigma) f(2 \rho_2 k) \} \right) \ge 0, \label{A7d4}
\end{equation}

\begin{equation}
A_8=\frac{5}{81} \left( f(6k) +6 f(3k) +8 f(-3k) + 2\Re \{ 4(\sigma-1) f(\rho_2 k) -2(2+\sigma) f(\rho_5 k) +(7+5\sigma) f(2 \rho_2 k) \} \right) \ge 0,
\end{equation}

\begin{equation}
A_9=\frac{2}{81} \left( f(6k) +15 f(3k) -10 f(-3k) + 2\Re \{ 10(\sigma-1) f(\rho_2 k) +(4\sigma-1) f(\rho_5 k) -5(1+2\sigma) f(2 \rho_2 k) \} \right) \le 0. \label{A9d4}
\end{equation}

All inequalities, numerically checked, hold for $0 \le k \le 0.616604$, where $f(-3k)$ is analytical, except Eq. (\ref{A1d4}), that holds for $0 \le k \le 0.540037$, and Eq. (\ref{A7d4}), that holds for $0 \le k \le 0.558283$. Again, above these points we get $A_1 < 0$ and $A_7 < 0$ respectively. As we shall show, the upper bound for the coupling constant lies in the domain of validity of Eqs. (\ref{A1d4})-(\ref{A9d4}).

\section{Application of the correlation inequalities}

In order to derive the coupling bounds for the area decay we use the same procedure as in the mean field derivation (section 2, equation (\ref{eq4})). Applying the same procedure to equations (\ref{SD3d}) and (\ref{SD4d}), we will prove the following statements.

Let $\beta $ be such that,

\vspace{0.5cm}
(a) $\textrm{  } 4(A_1 + A_2 + A_5) < 1$. Then, for $d = 3$
\begin{equation}
\langle W( C )\rangle  \leq  e^{-  \mid \ln( 4(A_1 + A_2 + A_5) \mid A} \label{eq29}.
\end{equation}

We numerically evaluate the condition $4(A_1 + A_2 + A_5) < 1$, resulting in $k < 0.440761$. It is worth to notice that inequalities (\ref{A1d3})-(\ref{A5d3}) are valid within the interval $0 \leq k \leq 0.440761$.

\vspace{0.5cm}
(b)$\textrm{  }6(A_1 + A_2 + A_6 + A_7 + A_8) < 1$. Then, for $d = 4$
\begin{equation}
\langle W( C )\rangle  \leq  e^{-  \mid \ln( 6(A_1 + A_2 + A_6 + A_7 + A_8) \mid A} \label{eq30}.
\end{equation}
Numerically evaluating the condition $6(A_1 + A_2 + A_6 + A_7 + A_8) < 1$ we obtain $k < 0.101220$, the upper bound for the area decay. All inequalities expressed by Eqs. (\ref{A1d4})-(\ref{A9d4}) hold in the domain $0 \le k \le 0.101220$.

\vspace{0.5cm}
Let us prove result (a). For b $ \epsilon$  C,
\begin{eqnarray}
\langle W(C)\rangle &=&
A_1 \sum_{i} \langle W(C) \chi^2_i \rangle 
+ A_2 \sum_{i<j} \langle W(C) \chi_i \chi_j \rangle 
+ A_3 \sum_{i<j<k} \langle W(C) \chi_i \chi^2_j \chi^2_k \rangle \nonumber \\
&+& A_4 \sum_{i<j<k<l} \langle W(C) \chi_i \chi_j \chi_k \chi^2_l \rangle  
+ A_5 \sum_{i<j<k<l}\langle W(C) \chi^2_i \chi^2_j \chi^2_k \chi^2_l\rangle \nonumber\\
&&\leq A_1 \sum_{i} \langle W(C) \chi^2_i \rangle 
+ A_2 \sum_{i<j} \langle W(C) \chi_i \chi_j \rangle 
+ A_5 \sum_{i<j<k<l}\langle W(C) \chi^2_i \chi^2_j \chi^2_k \chi^2_l\rangle
\label{eq27}
\end{eqnarray}

In the last inequality we have considered the fact that $A_3$ and $A_4$ are negatives, $A_1$, $A_2$, and $A_5$ are positives, and $\langle W(C)\chi_{P_i}\chi_{P_j}\chi_{P_k} \rangle $ is positive by Griffith's first inequality(\cite{Ganikhodjaev2007,Ganikhodjaev2010}). For the other functions we make use of the inequalities which are based on $\langle S_A^2 S_B\rangle   \leq  \langle S_B\rangle $, where $S_A^2 = \prod S_i^2$ and $S_B = \prod S_i$ (\cite{Ganikhodjaev2007,Ganikhodjaev2010,Braga1993,Braga1994}).   At each application of the equality we pass to an inequality by dropping the terms; after A steps we arrive at
\begin{equation}
\langle W( C )\rangle  \leq  (A_1 + A_2 + A_5)^A \text{(sum of } 4^A \text{ terms)} \label{eq16}        
\end{equation}  
where each term is less than one.

To prove result (b) for d=4 we proceed as previously for the result (a) for d=3, using equation (\ref{SD4d}). We can drop the terms in favor of an inequality. At each stage we have six terms from each $A_1$, $A_2$, $A_6$, $A_7$, and $A_8$ terms.

\section{Final Comments}
We have presented the application of the three-state generalized Ising  variables to the lattice gauge system, where the variables occupy the bonds of the lattice and are coupled by a four-spin interaction (a plaquette).  We have presented the derivation of correlation identities, which are exact in all dimensions and we made use of correlation inequalities for the generalized Potts model to obtain the upper bounds for the area decay of Wilson's loop. The coupling constants obtained for those bounds or behaviors are calculated  for d=3 and d=4. We obtain rigorous results that improve mean field calculations.

\vspace{1cm}

\end{document}